# Enhanced four-wave mixing with nonlinear plasmonic metasurfaces

Boyuan Jin and Christos Argyropoulos*

Department of Electrical and Computer Engineering, University of Nebraska-Lincoln, Lincoln, NE, 68588, USA

*christos.argyropoulos@unl.edu

*Plasmonic metasurfaces provide an effective way to increase the efficiency of several nonlinear processes while maintaining nanoscale dimensions. In this work, nonlinear metasurfaces based on film-coupled silver nanostripes loaded with Kerr nonlinear material are proposed to achieve efficient four-wave mixing (FWM). Highly localized plasmon resonances are formed in the nanogap between the metallic film and nanostripes. The local electric field is dramatically enhanced in this subwavelength nanoregion. These properties combined with the relaxed phase matching condition due to the ultrathin area lead to a giant FWM efficiency, which is enhanced by nineteen orders of magnitude compared to a bare silver screen. In addition, efficient visible and low-THz sources can be constructed based on the proposed nonlinear metasurfaces. The FWM generated coherent wave has a directional radiation pattern and its output power is relatively insensitive to the incident angles of the excitation sources. This radiated power can be further enhanced by increasing the excitation power. The dielectric nonlinear material placed in the nanogap is mainly responsible for the ultrastrong FWM response. Compact and efficient wave mixers and optical sources spanning different frequency ranges are envisioned to be designed based on the proposed nonlinear metasurface designs.*

# Introduction

Four-wave mixing (FWM) is a parametric nonlinear process with widespread applications in wavelength conversion, signal regeneration, switching, phase-sensitive amplification, nonlinear imaging, and entangled photon pair generation [1]-[6]. It is a byproduct of the Kerr nonlinear optical effect and has an almost instantaneous response on the order of femtoseconds [7]. For these reasons, it has great potentials towards the practical realization of ultrafast all-optical signal processing components. In addition, another advantage of FWM is that it can realize nonlinear frequency conversion and generation over a broad frequency range from visible to infrared (IR) and low-THz [8]-[9]. These new optical nonlinear sources will be compact and able to be easily integrated on photonic chips.

However, FWM processes have an extremely weak nature and suffer from very poor efficiencies [7]. Typical to third-order optical nonlinear effects, very high input intensities are required to excite these processes. This detrimental issue directly limits their practical applications. Recently, a new way to enhance optical nonlinear processes has been proposed. It is based on plasmonic nanostructures [10] instead of the usual all-dielectric nonlinear waveguides [7]. In these metallic plasmonic structures, the electromagnetic fields can be enhanced and confined in subwavelength volumes forming localized and surface plasmon polaritons [11]. The intrinsic nonlinear response of noble metals, as well as the enhanced field intensity due to the extreme confinement of electromagnetic radiation, serve as an excellent platform to boost several optical nonlinear processes [10]. For example, recently, it was reported that Kerr nonlinear effects and FWM can be enhanced with several plasmonic metamaterial configurations [12]-[15]. Additionally, the phase-matching requirements are relaxed in plasmonic nanostructures [16], which consists a major advantage for nonlinear applications and, in particular, for FWM processes [17].

Therefore, the key towards improving the efficiency of several optical nonlinear processes (FWM and others) is to increase the local field intensity along nonlinear materials, which are usually located in highly subwavelength regions [18]-[30]. Towards this end, it was recently presented that the generated FWM wave at a plane nonlinear metallic surface can be increased by three orders of magnitude due to the strong field enhancement obtained by surface plasmon polaritons (SPP) [31]. In another relevant study [32], a plasmonic grating was used to further enhance FWM. In this configuration, Fabry-Pérot (FP) interference modes were created inside the grating's metallic narrow grooves due to increased coupling with the incident light. Consequently, the FWM efficiency was further improved by two orders of magnitude because stronger confined fields were interacting with the nonlinearities of metal. Recently, it was theoretically shown that the FWM efficiency can be improved by one additional order of magnitude with the use of periodic plasmonic nanowires coupled to a metallic film [33]. This extra improvement resulted from the excitation of localized surface plasmon resonance (LSPR) modes characterized by ultrastrong fields confined at the nanogap between the nanowires and the substrate. However, the field confinement is restricted to a very small region in this configuration, which directly sets up an upper limit to the FWM efficiency. It was also shown that the radiated power outflow of the generated FWM wave has moderate values and is not directional. Hence, the practical FWM enhancement of this plasmonic configuration is still limited.

In this work, we propose an alternative nonlinear plasmonic configuration based on film-coupled silver (Ag) nanostripes with the geometry shown in Fig. 1. The height of the nanostripes is only 40nm and this ultrathin structure can be considered to be a planar optical metamaterial, alternatively called metasurface, operating in reflection [34]-[35]. Giant degenerate FWM efficiency is obtained with this nonlinear metasurface due to the remarkable field enhancement achieved at the nanogap of this structure. In addition, the FWM wavelength conversion can be achieved in highly subwavelength regions, i.e. the nanogap with 2nm thickness, which relaxes the phase-matching condition. Highly localized plasmon resonances are formed inside this nanoregion between the metallic film and nanostripes and the local field is dramatically enhanced. Kerr nonlinear material is loaded inside the nanogap and it is shown that the reflected FWM signal is mainly generated in this region. The nonlinearity of metal is also taken into account in all our calculations but it is found to weakly contribute to the FWM process. The quantitative nonlinear analysis based on full-wave simulations demonstrates giant FWM efficiency with an improvement in the radiative FWM power outflow by *up to nineteen orders* of magnitude compared to the FWM produced by a plain Ag film. In addition, the generated FWM wave has a highly directional radiation pattern and its enhanced radiative power outflow is relatively insensitive to the incident angles of the FWM input waves. These are ideal conditions to realize efficient generation of electromagnetic radiation over a broad frequency range. We explore several ways to match the incident wavelengths with the multiple resonant wavelengths of the proposed nonlinear metasurface in order to efficiently generate visible and low-THz electromagnetic radiation. The proposed designs can improve the currently immature development state of low-THz sources. Interestingly, the THz radiation intensity generated from the presented nonlinear metasurface can be increased monotonically as we increase the input radiation power. Novel compact and efficient visible, IR and THz optical sources and wave mixers can be designed based on the proposed ultrathin nonlinear device.

**Nonlinear Metasurface Geometry**

The geometry of the nonlinear plasmonic metasurface is shown in Fig. 1. Periodic nanostripes are coupled to a metallic film. All the metallic parts of the proposed metasurface designs are assumed to be made of silver. Note that similar structures have recently been experimentally verified to boost third-harmonic-generation (THG) [36]. The spacer layer placed between the nanostripes and the metallic film is filled with Kerr nonlinear material. It has an ultrathin thickness $g = 2nm$. The nanostripes are considered infinite along the z-direction, and their periodicity in the x-direction is $d = 250nm$. Each nanostripe has width $l = 40nm$ and height $h = 40nm$. To comply with experimental fabrication tolerances, the corners of their cross-sections are rounded with a radius $r = 2nm$ [36]. The thickness of the metallic film (H=80nm) is chosen to be larger than silver's skin depth at optical and THz frequencies in order for the transmission to be equal to zero.

Two waves with different angular frequencies $\omega_1$ and $\omega_2$ are impinging on the metasurface along the x-y plane. Their incident angles are $\theta_1$ and $\theta_2$, respectively. A new wave with frequency $\omega_3$ is generated via the degenerate FWM nonlinear mechanism following the relationship: $\omega_3 = 2\omega_1 - \omega_2$ [37]. The incident waves are always transverse magnetic (TM) polarized and their magnetic field direction is along the z-axis. The proposed metasurface is expected to exhibit polarization dependence but its response is relatively independent to the angle of incidence of the impinging waves, as it will be shown later. The proposed design can become polarization independent, if the nanostripes are replaced by their three-dimensional (3D) counterparts, an array of nanocube resonators [38]-[45]. The physics of these two plasmonic systems are very similar except that the latter one can work for both polarizations, i.e., it is polarization insensitive. Hence, both systems can enhance several nonlinear effects and other photodynamic processes, such as spontaneous emission rates [38]-[45].

**Results and Discussion**

Linear simulations are performed when we set-up the nonlinear part of metal and dielectric permittivities equal to zero $\chi_d^{(3)} = \chi_{Ag}^{(3)} = 0$. The metasurface shown in Fig. 1 is illuminated with two normal incident TM polarized plane waves. The dimensions are given in the previous section. The reflectance is plotted in Fig. 2 as a function of the excitation wavelength, and the fundamental resonant wavelength is computed to be located at $833nm$. The reflectance decreases significantly at this resonant frequency point and a localized plasmon resonance is formed. The impinging radiation is trapped in the spacer layer, leading to enhanced fields inside the nanogap between the film and the nanostripes with a standing wave FP distribution [36], [46]. The amplitude of the resonant electric field enhancement distribution has been computed and is shown in the inset of Fig. 1. The local maximum of the field enhancement can reach large values on the order of $|E/E_0|_{max} = 250$ at the resonance, where $E_0$ is the electric field amplitude of the incident wave. The maximum field enhancement occurs near the edges of the nanostripe, which is consistent to a FP resonance field distribution. Both average and maximum local field enhancements in the nanogap are plotted in Fig. 2 (dotted and solid red lines, respectively). By comparing the values of these two parameters, the degree of uniformity of the electric field inside the nanogap can be derived. The strong field enhancement confined in an extremely subwavelength region $(g = 2nm)$ is ideal condition to boost several nonlinear processes, such as FWM, second and third harmonic generation. Note that the resonant frequency response of the proposed metasurface can be tuned throughout the visible and IR by just varying its geometric parameters [42]. The resonant wavelength will change when we vary the thickness of the spacer layer $g$, the width of the nanostripes $l$, and the rounded corners (radius $r$) of the nanostripe cross-sections. On the contrary, the field enhancement and the resonant frequency will be weakly impacted by the periodicity $d$ and the incident angles $\theta_1$ and $\theta_2$. The field enhancement is maximum at normal incidence and slightly decreases towards grazing angles [43]. This will moderately affect the FWM efficiency at grazing angles, as it will be shown later in the manuscript.

To utilize the strong resonant field enhancement effect, the incident waves are assumed to be monochromatic with input wavelengths $\lambda_1 = 833 nm$ and $\lambda_1 = 845 nm$. They have equal optical intensities $I_1 = I_2 = 3 MW/cm^2$, which are much lower compared to previous relevant FWM works [33]. These values are well below the damage threshold of silver or other metal and dielectric materials used to construct the proposed nonlinear metasurface. The generated FWM wave will have a wavelength of $\lambda_3 = 821$ nm, following the frequency mixing relation presented before. The wavelengths of both incident waves and the generated FWM wave are all located close to the fundamental resonance $(833 nm)$ of this metasurface. As a result, the induced fields at the nanogap of each incident and generated wave are drastically enhanced at these frequency points.

In experimental set-ups, the measured power of the generated FWM wave will be equal to the power flow radiated through the boundaries of the current simulation domain [33]. Therefore, the FWM efficiency can be described in a more practical way by computing the power outflow of the generated FWM wave. The FWM process is relatively insensitive to both incident angles of the input waves for this particular plasmonic metasurface, as it will be shown in the next section.

Hence, for simplicity, $\theta_2$ is kept constant and equal to zero $(\theta_2 = 0°)$, while $\theta_1$ varies from $-90^o$ to $90^o$. The FWM power outflow is computed and is found to be symmetric with respect to $\theta_1 = 0°$. The result is shown in Fig. 3 (solid line). The maximum power outflow is $2.2 \times 10^5$ W/m at $\theta_1 = 0°$. We have also computed the FWM power outflow in two other structures: a bare silver film (dashed line in Fig. 3) and a silver film with a 2nm thick nonlinear dielectric layer on top of it (dotted line in Fig. 3). The dielectric layer has the same linear and nonlinear properties with the dielectric loaded in the metasurface nanogap. The two small peaks at $\theta_1 = \pm 30°$ in the dashed and dotted lines of Fig. 3 are due to the surface plasmon resonant enhancement effects generated from the silver layer [31]. Interestingly, it can be seen that the power outflow is *increased with the proposed metasurface by nineteen and sixteen orders of magnitude*, respectively, compared to the other two structures. This giant FWM efficiency enhancement can be triggered with low input intensities. It will facilitate the efficient excitation of FWM nonlinear signals by using nanoscale devices.

In addition to high efficiency, the FWM radiated wave produced by the proposed nonlinear metasurface is homogeneous and relatively insensitive to the angles of incident waves. This can be partially deduced by Fig. 3, due to the flat curve of the produced FWM wave for different incident wave angles $(\theta_1)$. Similar insensitive angle operation is expected for the other incident wave impinging with an angle $\theta_2$. To further illustrate this interesting effect, the distribution of the FWM power outflow as a function of the excitation angles of both incident waves ($\theta_1$ and $\theta_2$) is computed and plotted in Figs. 4(a) and (b) for the proposed nonlinear metasurface and the bare Ag film, respectively. The FWM efficiency of the proposed nonlinear metasurface is extremely high and uniformly distributed within a wide range of both incident waves excitation angles. It declines only when the angles approach grazing incidence [Fig. 4(a)]. On the contrary, the FWM distribution is very sensitive to both excitation angles in the bare silver film case. The efficiency is very small and is dramatically altered with the direction of the incident waves [Fig. 4(b)]. The peaks in the FWM distribution are attributed to the excited surface plasmon polaritons propagating on the surface of the silver layer [31].

As indicated before, the resonant wavelengths can be adjusted just by changing the dimensions of the metasurface. Figure 5 shows the reflectance of the linear metasurface when the width of the nanostripe is increased to $l = 70nm$ and all the remaining dimensions are kept the same with before. In this design, the fundamental resonance is red-shifted from 833nm to 1206nm. In addition, higher-order resonances are supported by this metasurface, which are always located at lower wavelengths compared to the fundamental resonance (Fig. 5). The maximum and average electric field enhancements are also plotted in Fig. 5 (red solid and dotted lines, respectively). The field is enhanced at both fundamental and higher-order resonances located at $\lambda = 1206nm$ and $\lambda = 620nm$, respectively. However, the local maximum of the field enhancement in the nanogap is $|E/E_0|_{max} = 60$ at the higher-order resonance (Fig. 5), which is relatively smaller compared to the field enhancement at the fundamental resonance $\left(|E/E_0|_{max} = 185\right)$. The average value of the field enhancement is also smaller $\left(|E/E_0|_{avg} = 30\right)$ at the higher-order resonance compared to the value at the fundamental resonance $\left(|E/E_0|_{avg} = 84\right)$. The field enhancement distribution at the higher-order resonance is shown in the inset of Fig. 5. In this case, there are 4 antinodes in the electric field distribution which appear between the nanostripe and the metallic layer, while only 2 antinodes emerge at the fundamental resonance (see inset in Fig. 1). This is a typical higher-order FP resonant wave distribution.

According to the relationship $\omega_3 = 2\omega_1 - \omega_2$, the frequencies $\omega_1$, $\omega_2$ of the two monochromatic incident waves should be largely separated in order to produce a FWM signal $(\omega_3)$ at low-THz frequencies. The efficiency of the FWM generated THz wave can be enhanced by matching the input wavelengths to more than one resonances of the proposed metasurface. Towards this end, two monochromatic plane waves with $\lambda_1 = 1206 nm$ and $\lambda_2 = 620 nm$ are launched to the proposed metasurface with the dimensions mentioned before. The incident wavelengths coincide with the fundamental and higher-order resonances, as they were computed in Fig. 5. The induced fields at the nanogap are enhanced for both incident wave frequencies leading to an efficient FWM process.

The power outflow of the generated FWM wave is again computed by nonlinear simulations based on COMSOL and all the nonlinear material parameters are the same with the previous example. It is plotted in Fig. 6, as a function of the incident angle $\theta_1$, for the proposed nonlinear metasurface (solid line), the bare silver film (dashed line) and the silver film covered by a 2nm thick nonlinear dielectric layer (dotted line). The FWM power outflow of the proposed nonlinear metasurface is i*ncreased by fifteen and thirteen orders of magnitude* compared to the other configurations. Furthermore, it is noteworthy that now the wavelength of the FWM generated wave is 22μm, corresponding to an approximate frequency of 13.5THz. Hence, the proposed nonlinear metasurface can provide a new approach to realize an efficient all-optical coherent THz source based only on ultrafast third-order optical nonlinearities. Recently, in an analogous way, it was demonstrated that an efficient THz source can be implemented by second-order nonlinear processes using the surface plasmon polaritons induced along the surface of graphene [47].

The maximum power outflow of the proposed structure is $5.46 \times 10^{-5}$ W/m at $\theta_1 = 0°$, which is lower compared to the previous FWM metasurface design, where the incident waves were located close to the fundamental resonance. This is due to the lower field enhancement in the nanogap of this configuration. The maximum field enhancement at the fundamental resonance is decreased to 185 when the nanostripe width is increased to $l = 70 nm$. In addition, the field enhancement at the higher-order resonance is always smaller compared to the fundamental resonance. Moreover, the FWM generated THz wave cannot be enhanced by the resonant effect of this structure because it is located in larger wavelengths, where the metasurface is not anymore resonant and a very small portion of the field is coupled inside the nanogap. There is a large frequency span among the wavelengths $\lambda_1$, $\lambda_2$, and $\lambda_3$ used in the current FWM process. However, the FWM output radiation power is still relatively high and can be further increased in real-time depending on the input intensities. In principle, arbitrary large THz radiation intensity can be generated with the proposed metasurface with values directly proportional to the input intensities. Its THz response is only limited by the damage threshold of the materials used to construct the proposed device. Note that phase mismatch is not an issue in this case, since the FWM process takes place in a highly subwavelength region inside the nanogap. This is in accordance to previous studies, where it was shown that the phase-matching condition can be relaxed in ultrathin nonlinear metasurfaces [16], [48]-[50].

While the FWM process is excited by visible or near-IR incident waves, the proposed nonlinear metasurface can operate as visible, IR or THz source. Coherent radiation is generated due to the ultrafast coherent nonlinear interactions between the incident light beams. However, the directivity of this FWM radiation source should also be computed in order to fully evaluate its practical potential. To this end, the far-field radiation patterns of the FWM generated signals for both previously presented nonlinear metasurfaces are computed and plotted in Fig. 7. The radiation patterns are plotted along the x-y plane, where the 0° angle represents the direction parallel to the silver film surface. For both visible and THz FWM-based sources, the intensity of the radiated wave is mainly concentrated along the y-axis, which is the direction perpendicular to the nonlinear metasurface. Directional FWM radiation patterns are obtained for both nonlinear metasurface designs. The thickness of the silver film is much larger than the skin depth and, as a result, almost no power of the generated FWM wave penetrates through the silver film. The nonlinear metasurface works only in reflection and the radiated power is approximately equal to zero in the angular range from 180° to 360°, as it is shown in Fig. 7. We also plot in Fig. 7 the radiation patterns for different excitation angles. The maximum radiation power drops when the excitation angles are increased, which is consistent with the FWM distribution results shown in Figs. 3 and 6, respectively. However, the directivity shape of the radiated power is independent of the excitation angles and always a directive spatially-coherent emission is obtained.

It is favorable in the design of efficient electromagnetic radiation sources to be able to control, tune and increase the output radiation power of the generated wave. This can be easily achieved with the current FWM configurations by varying the incident power of the excitation waves. Figure 8 demonstrates the effect of incident intensities $P_1$ and $P_2$ on the generated FWM power for both visible [Fig. 8(a)] and THz [Fig. 8(b)] source configurations. The FWM power outflow is approximately a square function of $P_1$ and a linear function of $P_2$. The two incident waves have different influences on the power of the generated FWM wave, as expected by Eq. (1). Silver will not be damaged, i.e. melt, at these low input pump intensities and the generated output power can be arbitrary increased as long as the damage threshold of the used materials is not reached. Therefore, derived from Fig. 8, increasing the input pump power is an effective way to enhance the FWM radiated power.

Finally, in all our nonlinear metasurface simulations, the FWM process occurred in both the metallic parts and the dielectric spacer layer. To find out which part dominates in the FWM mechanism, we calculated the FWM efficiency when we ignore the nonlinearity of either the metallic or dielectric material. The first metasurface design, with FWM generated power results shown before in Fig. 3, is employed for this comparison. When the nonlinearity of silver is not present $\left(\chi_{Ag}^{(3)}=0\right)$, the FWM output power is almost the same with Fig. 3, as it is shown in Fig. 9. On the contrary, if the nonlinearity of the dielectric spacer is ignored $\left(\chi_{d}^{(3)}=0\right)$ and the metallic parts provide the nonlinear response, the power outflow is dramatically decreased by eleven orders of magnitude. Therefore, it can be concluded that the FWM is mainly generated by the nonlinear dielectric spacer layer. This is consistent with the results shown in the insets of Figs. 1 and 5, where the fields have been found to be primarily enhanced between the nanostripes and the silver film, mainly inside the nanogap region. The fields cannot penetrate the metallic parts of the nonlinear plasmonic metasurface and, as a result, cannot interact with the nonlinearity of metal. Hence, the FWM signal generated by the nonlinear dielectric placed in the spacer layer is dominant. Note that if the nonlinear dielectric placed in the spacer layer is changed to materials with higher nonlinear coefficients, such as organic polymers [51], the radiation power of the generated FWM wave will be further increased.

**Conclusions**

Efficient FWM in nanoscale regions requires enormous input optical intensities to be excited, which makes its practical implementation very challenging. In this work, we demonstrated that nonlinear metasurfaces based on film-coupled silver nanostripes can dramatically enhance FWM effects in the nanoscale. Due to the strong localized plasmon resonance at these structures, the optical field is dramatically enhanced and confined in the nanogap region between the metallic film and the nanostripes. This field intensity enhancement along the nonlinear material in conjunction with the relaxed FWM phase-matching conditions due to the metasurface ultrathin thickness led to giant improvement in the FWM efficiency. By decorating a bare silver film with an array of silver nanorods, the efficiency of FWM and other nonlinear processes can be enhanced by many orders of magnitude. In particular, when the wavelengths of both incident waves are close to the fundamental resonance, the power outflow of the FWM generated wave can be improved by approximately *nineteen orders of magnitude* compared to a bare silver film. Besides, when the incident wavelengths are matched with multiple resonances of the metasurface, an efficient THz source can be realized, which can be nonlinearly excited by visible and near-IR radiation. In this case, the FWM generated power outflow is increased by *sixteen orders of magnitude* compared to a bare silver film. Furthermore, the coherent FWM generated wave has a directional radiation pattern and the strong FWM efficiency is relatively homogeneous and insensitive to the incident angles of the excitation waves. This efficiency can be further improved by increasing the input intensities of the pump waves or by choosing materials placed in the spacer layer with higher nonlinear coefficients. To conclude, optical losses weakly affect the nonlinear FWM performance in the configuration under study and a giant enhancement in the effective dielectric nonlinear coefficient is obtained by just decorating the metal substrate with metal nanostripes. An alternative robust way is proposed to generate high intensity directional low-THz radiation, which is a very challenging task. The proposed nonlinear plasmonic metasurface provides a mean to introduce extremely large nonlinear enhancements that can be useful in generating entangled photon pairs or other wave mixing procedures, allowing a huge reduction in the intensity of the required input sources.

**Methods**

The FWM process will induce a nonlinear polarization given by [7]:

$$P_{NL} = \varepsilon_0 \chi^{(3)}(\omega_3; \omega_1, \omega_1, -\omega_2) E_1 E_1 E_2^*, \tag{1}$$

where $\chi^{(3)}$ is the third-order nonlinear susceptibility of both dielectric and metallic materials used in the design of this metasurface. The dielectric spacer layer is made of Kerr nonlinear material $\varepsilon = \varepsilon_L + \chi^{(3)} |E|^2$ with a linear permittivity value $\varepsilon_L = 2.2$ and a third-order nonlinear susceptibility $\chi_d^{(3)} = 4.4 \times 10^{-18} m^2/V^2$, typical values of polymers and semiconductors [7], [40]. The silver metallic parts of the metasurface are also characterized by a Kerr nonlinear permittivity with $\chi_{Ag}^{(3)} = 9.3 \times 10^{-20} m^2/V^2$ [7]. Their complex linear permittivity is obtained from experimental values [52].

The wave equation, including the nonlinear polarization, is equal to [53]:

$$[\nabla \times (\nabla \times) - \varepsilon(\omega_3) \omega_3^2 / c^2] E_3 = \omega_3^2 \mu_0 P_{NL}. \tag{2}$$

By solving this nonlinear wave equation with the electromagnetic solver of COMSOL Multiphysics, a commercial software based on the finite element method (FEM), the FWM properties of the nonlinear plasmonic metasurface are computed. Our proposed structure is modeled as a two dimensional system, where the nanostripes are assumed to be infinitely long in the z-direction. This assumption is reasonable because the length of the nanostripes is much larger compared to their cross-section dimensions. Three coupled wave equations at frequencies $\omega_1$, $\omega_2$, and $\omega_3$, respectively, are solved in the frequency domain, corresponding to incident and generated waves. Our proposed structure is periodic at the horizontal x-direction and periodic boundary conditions are applied at the two vertical boundaries enclosing the unit cell of one nanostripe. Thus, only one nanostripe (i.e. one period) needs to be included in the simulation domain, which effectively accelerates the nonlinear calculations.

## Acknowledgments

This work was partially supported by the Office of Research and Economic Development at University of Nebraska-Lincoln and NSF Nebraska MRSEC.

**Figures**

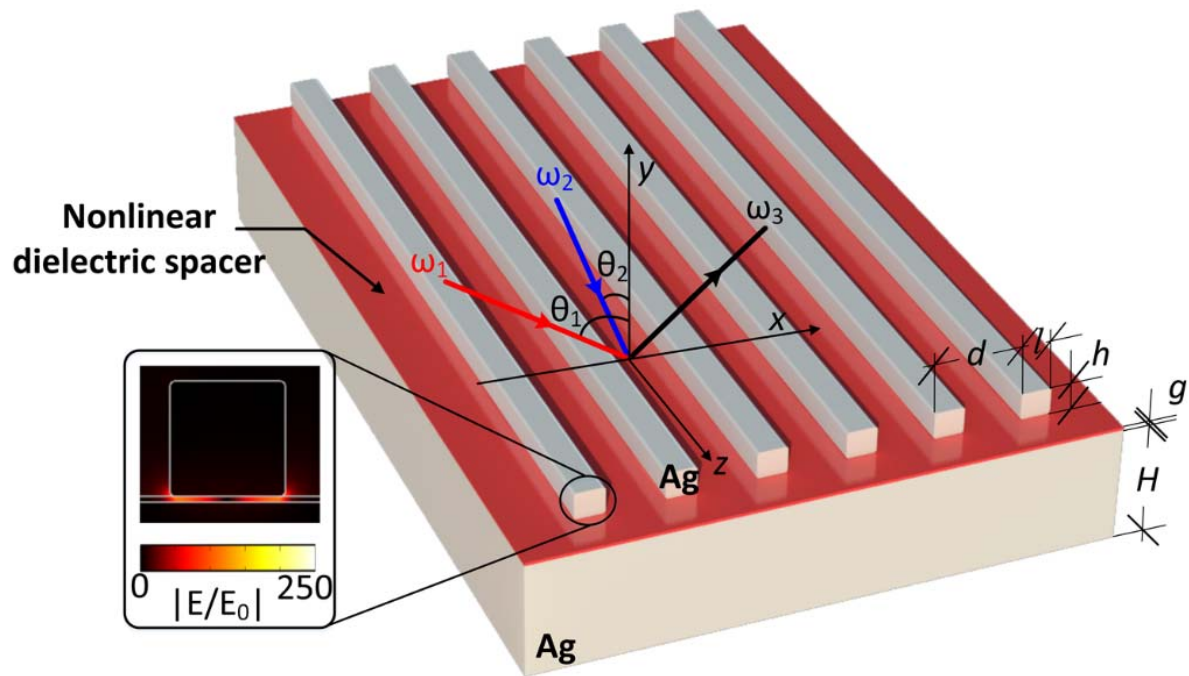

Figure 1 –Schematic illustration of the plasmonic metasurface based on silver nanostripes coupled to a silver film. Inset: Distribution of the field enhancement at the fundamental resonance (λ=833nm).

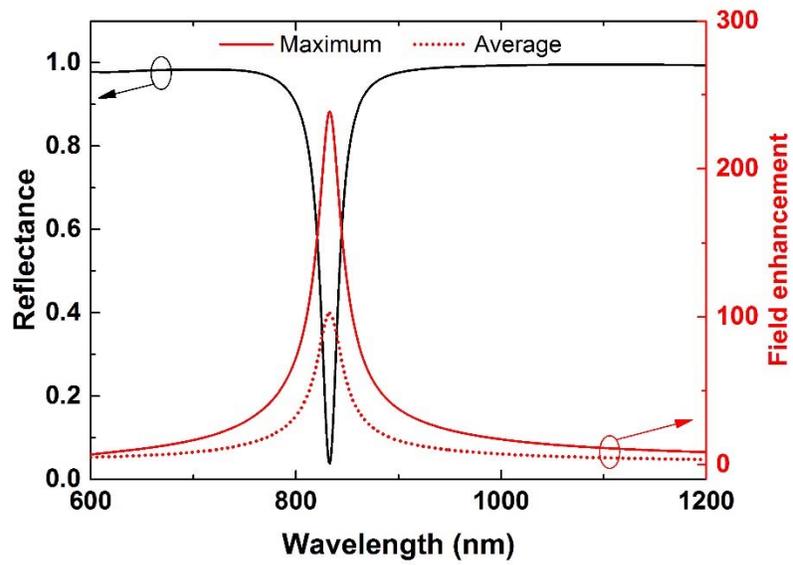

Figure 2 –Reflectance (black line) and field enhancement (red lines) distributions of the linear metasurface versus the incident wavelength. The red solid and dotted lines depict the local maximum and the spatially averaged field enhancement, respectively.

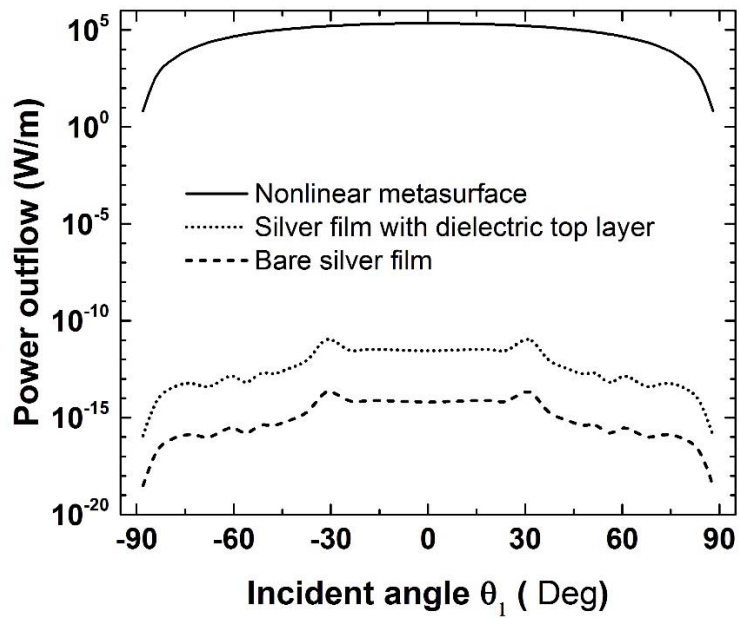

Figure 3 –Power outflow of the generated FWM wave as a function of different excitation angles for the nonlinear metasurface (solid line), a silver film with a nonlinear dielectric top layer (dotted line) and a bare silver film (dashed line). The incident wavelengths are $\lambda_1 = 833nm$ and $\lambda_2 = 845nm$ and the wavelength of the generated FWM wave is $\lambda_3 = 821nm$. The generated FWM power outflow of the nonlinear metasurface is enhanced by nineteen orders of magnitude compared to a bare silver film.

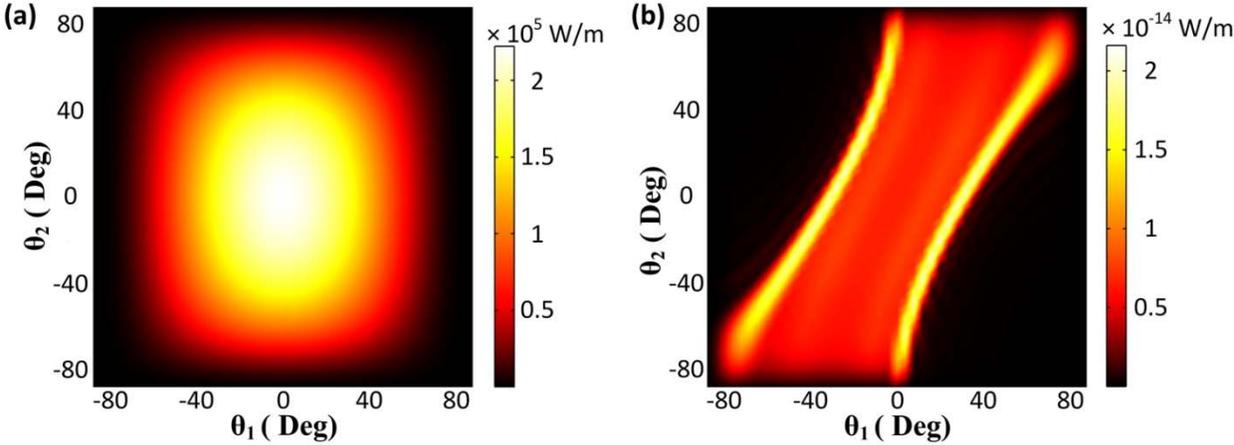

Figure 4 –Power outflow of the generated FWM wave as a function of the incident angles of both excitations for (a) the nonlinear metasurface and (b) the bare silver film. The generated FWM power outflow of the nonlinear metasurface is enhanced, homogeneous and relatively insensitive to the incident angles compared to a bare silver film.

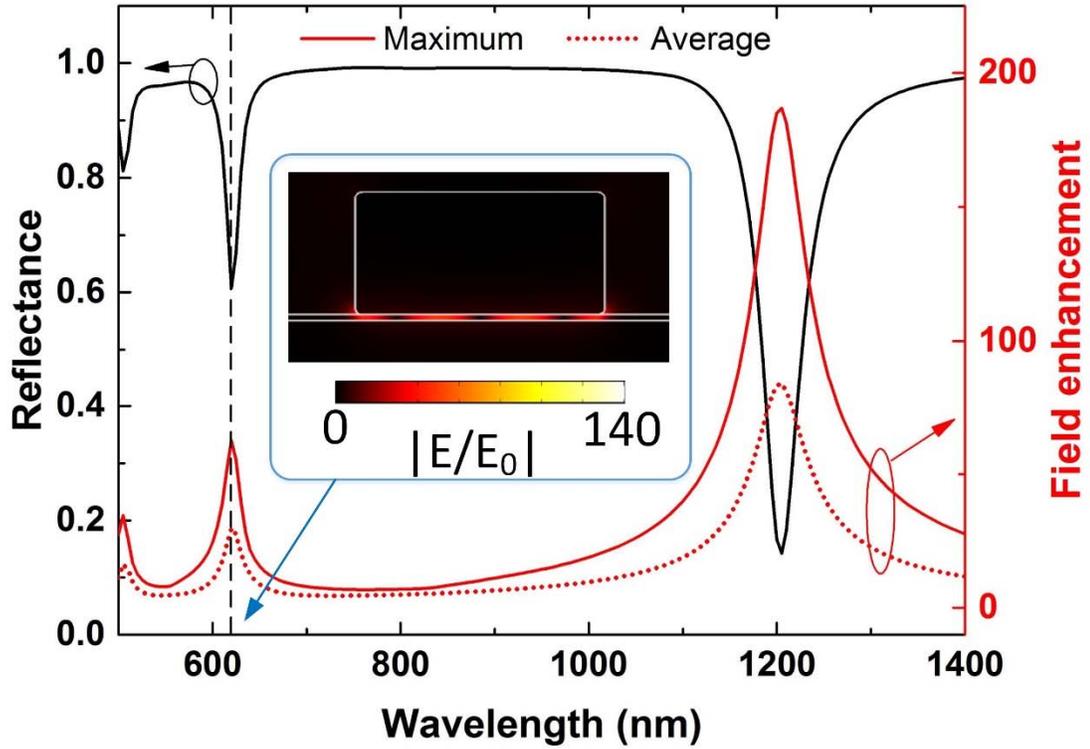

Figure 5 – Reflectance (black line) and field enhancement (red lines) distributions of the linear metasurface with nanostripes width $l = 70$ nm. The red solid and dotted lines depict the local maximum and the spatially averaged field enhancement, respectively. Inset: Distribution of the field enhancement at the higher-order resonance ($\lambda$=620nm).

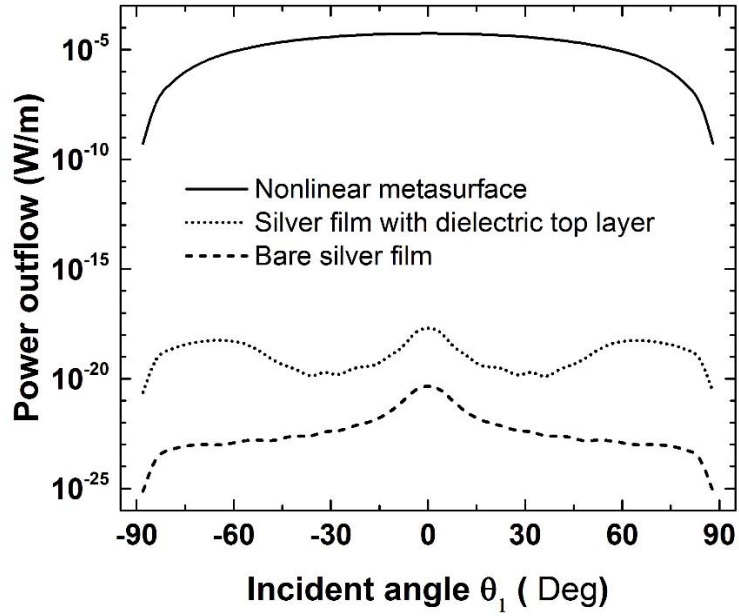

Figure 6 –Power outflow of the generated 13.5THz FWM wave as a function of different excitation angles for the nonlinear metasurface (solid line), a silver film with a nonlinear dielectric top layer (dotted line) and a bare silver film (dashed line). The incident wavelengths of the pump waves are $\lambda_1 = 1206nm$ and $\lambda_2 = 620nm$ and the wavelength of the generated low-THz FWM wave is $\lambda_3 = 22\mu m$. The power outflow of the generated 13.5THz FWM wave is enhanced by fifteen orders of magnitude compared to a bare silver film.

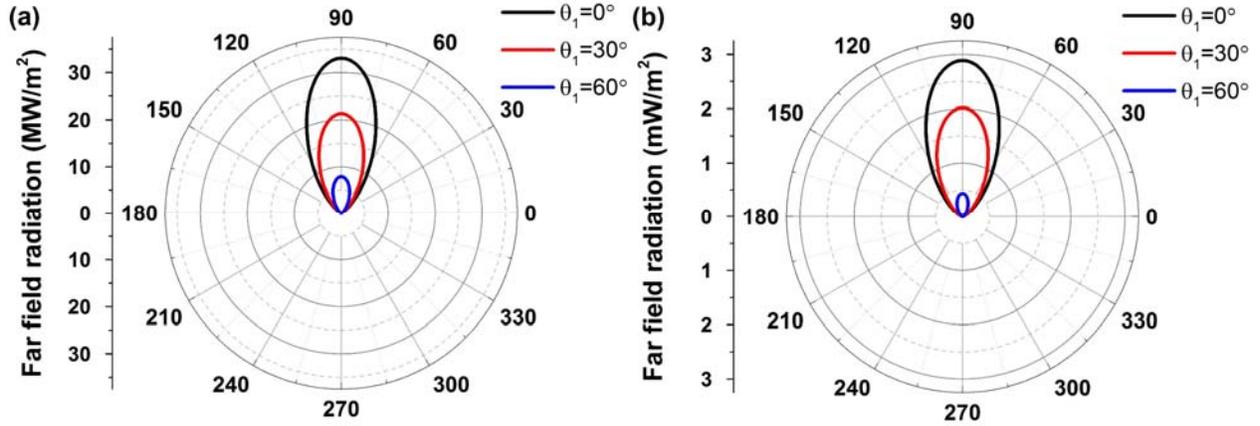

Figure 7 –Far field FWM radiation patterns of the nonlinear metasurfaces for incident pump waves (a) $\lambda_1 = 833nm$ and $\lambda_2 = 845nm$ (Fig. 1 design), and (b) $\lambda_1 = 1206nm$ and $\lambda_2 = 620nm$ (Fig. 5 design) impinging with different incident angles. Directional FWM radiation patterns are obtained for both nonlinear metasurface designs.

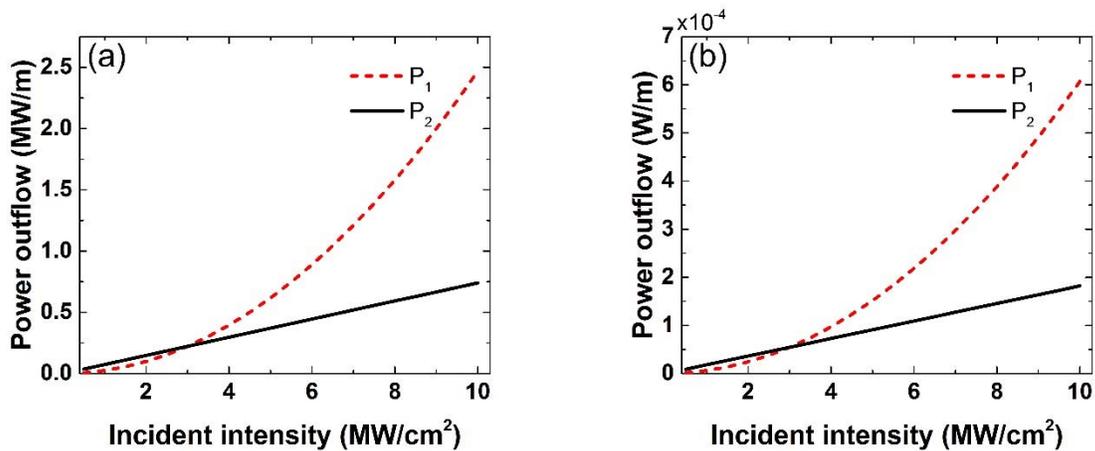

Figure 8 –The effect of the incident pump intensities $P_1$ and $P_2$ on the power outflow of the generated FWM wave. The incident pump wavelengths are (a) $\lambda_1 = 833nm$ and $\lambda_2 = 845nm$ (Fig. 1 design), and (b) $\lambda_1 = 1206nm$ and $\lambda_2 = 620nm$ (Fig. 5 design), respectively. The power

outflow of the generated FWM waves can be controlled and increased by varying the power of the incident waves.

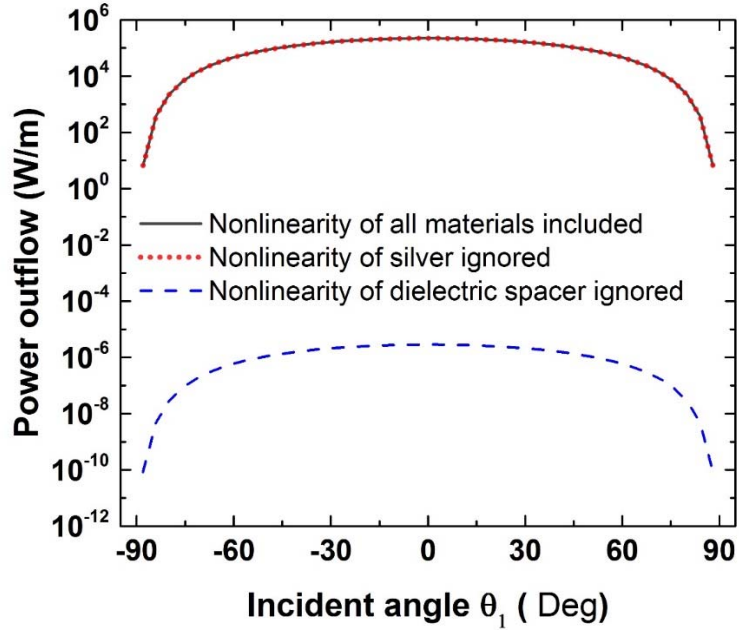

Figure 9 –The FWM power outflow of the nonlinear metasurface (Fig. 1 design) when the nonlinearities of all materials are included (black solid line), the nonlinearities of only the silver parts are ignored (red dotted line) and the nonlinearity of only the dielectric spacer layer is ignored (blue dashed line). It can be concluded that the FWM is mainly generated by the nonlinear dielectric spacer layer.